\documentclass[12pt,a4paper]{article}
\usepackage[pdfstartview=FitH,colorlinks=true,linkcolor=blue,anchorcolor=red,citecolor=magenta,urlcolor=blue]{hyperref}
\usepackage[english]{babel}
\usepackage{amsmath,amssymb,titling,authblk}
\usepackage{slashed}
\usepackage{amsmath}
\usepackage{amscd}
\usepackage[normalem]{ulem}
\usepackage{appendix}
\usepackage{bbold}
\usepackage{bm}

\usepackage{pdflscape}
\usepackage{yfonts}
\usepackage{amscd}
\usepackage{epsfig}
 \usepackage{microtype}

\usepackage{cite}

\allowdisplaybreaks[4]
\numberwithin{equation}{section}

\usepackage {color}
\usepackage{bbold}
\usepackage{braket}
\usepackage{bbm}

\definecolor{verde}{cmyk}{.83,.21,1,.08}

\advance\hoffset by -1.1truecm
\advance\voffset by -1.0truecm
\newcommand{\be}{\begin{equation}}
\newcommand{\ee}{\end{equation}}
\newcommand{\bea}{\begin{eqnarray}}
\newcommand{\eea}{\end{eqnarray}}

\newcommand{\del}{\partial}
\newcommand{\dd}{\mathrm d}
\newcommand{\ii}{\mathrm i}
\newcommand{\e}{\mathrm e}
\usepackage{mathrsfs}
\usepackage{authblk}

\numberwithin{equation}{section}
\usepackage{float}
\restylefloat{figure}
\newcounter{appendice}

\usepackage[T1]{fontenc}
\usepackage{lmodern}

\begin{document}

\setlength{\droptitle}{-6pc}

\title{Localization and observers in $\varrho$-Minkowski spacetime}

\renewcommand\Affilfont{\itshape}
\setlength{\affilsep}{1.5em}

\author[1,2,3]{Fedele Lizzi\thanks{fedele.lizzi@na.infn.it, fedele.lizzi@unina.it}}
\author[1]{Luca Scala\thanks{luca.scala.1997@gmail.com}}
\author[1,2]{Patrizia Vitale\thanks{patrizia.vitale@na.infn.it}}
\affil[1]{Dipartimento di Fisica ``Ettore Pancini'', Universit\`{a} di Napoli {\sl Federico~II}, Napoli, Italy}
\affil[2]{INFN, Sezione di Napoli, Italy}
\affil[3]{Departament de F\'{\i}sica Qu\`antica i Astrof\'{\i}sica and Institut de C\'{\i}encies del Cosmos (ICCUB),
Universitat de Barcelona, Barcelona, Spain}

\date{}

\maketitle

\vspace{-2cm}

\begin{abstract}\noindent
We consider the $\varrho$-Minkowski spacetime, a  model with linear noncommutativity involving the time and the azimuthal angle. We study its quantum symmetries, the $\varrho$-Poincar\'e\ quantum group, and  analyze the concepts of localizability and quantum observers.
 \end{abstract}

\newpage

\section{Introduction} \label{intro}
In this work we investigate the basic features of localizability in a specific model of noncommutative spacetime, $\varrho$-Minkowski, based on a previous analysis~\cite{lizzimercati, lizzimercati2} which has been carried out for the more famous $\kappa$-Minkowski spacetime. 

The interest for quantum, or noncommutative, spacetimes is deeply motivated by the search for a consistent theory of quantum gravity, valid at Planck scale, of which quantum spacetimes should represent a signature at lower energies. Noncommutative spacetimes, whose symmetry groups are deformations of the Poincar\'e\ group, are therefore the natural candidates to be investigated. 

The space we investigate, $\varrho$-Minkowski, has a noncommutativity of the angular type, involving time and the azimuthal angle. Time has a quantized discrete spectrum. Noncommutativity in the coordinates implies the presence of uncertainty relations among time and angle. A perfectly localized state in time is totally spread in $\varphi$, and vice versa, angle localization increases the time measurement uncertainty.

The presence of a noncommutative spacetime, invariant under a quantum group, implies that also observers, reference frames, have to be quantized, and this is the important feature of this activity. Observers are quantum objects as well. We have collected observations and definitions about states, observers and observables in the Appendix~\ref{AppA}.

We start in Section \ref{rhospacetime} by defining the commutation relations of the $\varrho$-Minkowski model and showing explicitly its angular nature. In Section \ref{rhovskappa}, we obtain the $\kappa$-Poincar\'e\ quantum group in order to illustrate the differences between the $\varrho$ and $\kappa$ deformations and explain the approach carried on for the $\varrho$ case by means of comparison. Once having obtained the $\varrho$-Poincar\'e\ quantum group as the symmetry group of our model, we tackle in Section \ref{section3} the problem of localizability in the $\varrho$-Minkowski spacetime, comparing the results with the known ones for the $\kappa$ case. To this, we realise both the spacetime observables and the quantum group generators as operators and represent them on a suitable Hilbert space. Section \ref{conc}, dealing with conclusions and perspectives, closes the work.
In appendix \ref{AppA}, as we mentioned, the notions of states, observables and observers employed trough the paper are formally stated, while
appendix~\ref{App1.4} is devoted to recalling the classical $r$-matrix deformation method to obtain the Poincar\'e\ quantum groups.

\section{The \texorpdfstring{$\varrho$}{}-Minkowski spacetime} \label{rhospacetime}
The $\varrho$-Minkowski spacetime  is characterized by the following commutation relations of the angular type:
\bea
{}[x^0,x^1]&=&\phantom{-}\ii\varrho x^2\nonumber\\
{}[x^0,x^2]&=&-\ii\varrho x^1 \label{rhoMincommrel}
\eea
all other commutators being zero. In particular, $x^3$ is central; it commutes with all coordinates. The $\varrho$ parameter has the dimension of a length, and it is often identified with the Planck length, the scale at which quantum gravity effects are expected to be manifest.

These relations are a part of a larger family of ``Lie-algebra-type'' commutation relations, of which the most famous case is $\kappa$-Minkowski spacetime defined by\footnote{We use the standard convention for which the latin indices $i,j\ldots$ go from 1 to 3, while the greek ones $\mu,\nu\ldots$ go from 0 to 3.}
\be
[x^0,x^i]=\ii\lambda x^i \, ,\label{kappaMincommrel}
\ee 
again all other commutators vanishing. The $\lambda$ parameter with the dimension of a length is sometimes expressed as $\frac1\kappa$, hence the name of the model.

The Lie algebra~\eqref{rhoMincommrel} has the structure of  the  Euclidean algebra in $2+1$ dimensions; it goes back to at least~\cite{Gutt} (also see~\cite{selene,PacholVitale}). In the context
of twisted symmetries it was analyzed by Lukierski and Woronowicz in~\cite{Lukierski:2005fc}. 
In ~\cite{AmelinoCamelia:2011gy}  it was discussed in relation with  the principle of relative
locality~\cite{AmelinoCamelia:2011bm}. This kind of noncommutative spacetime might have concrete physical relevance~\cite{Ciric:2017rnf, Ciric:2019urb, Ciric:2019uab} and 
phenomenological/observational consequences~\cite{Amelino-Camelia:2017pne}. In~\cite{DimitrijevicCiric:2018blz},  a field theory on this space has been built; in the same paper, a different physical identification of the noncommuting variables has been also considered, with time  a commuting coordinate. The latter has been studied in ~\cite{Kurkov:2021kxa} in the context of Poisson gauge models and in~\cite{Gubitosi:2021itz} in the context of double quantization\footnote{In order to distinguish the two spacetimes, the one with commutative time has been named  $\lambda$-Minkowski spacetime}. 

The commutation relations~\eqref{rhoMincommrel} give rise to two nontrivial uncertainty relations:
\begin{align}
\Delta x^0 \Delta x^1 &\geq \frac\varrho2 \left|\langle x^2\rangle\right|\nonumber\\
\Delta x^0 \Delta x^2 &\geq \frac\varrho2 \left|\langle x^1\rangle\right|. \label{3.24}
\end{align}
This implies that in this  kind of noncommutative spacetime, sharp localization of event operators is not always possible.  Note that, by the centrality of $x^3$, this coordinate can be determined with absolute precision.

A realization of $\varrho$-Minkowski is given by~\cite{lizzivitale}
\begin{align}
x^i \psi (x) &=x^i \psi (x),\nonumber\\
x^0 \psi (x) &=-i\varrho (x^1 \partial_2-x^2\partial_1) \psi(x), \label{3.25}
\end{align}
with $x^i$ a complete set of observables on the Hilbert space $L^2(\mathbb{R}^3)$, $x^0$ a self-adjoint operator on $L^2(\mathbb{R}^3)$ acting like an angular momentum along the 3-axis, and $\psi(x)$ a state in the Hilbert space. 

We can choose a more convenient way of writing commutators and uncertainty relations, given by the fact that the $\varrho$ deformation is of angular nature. We therefore use cylindrical coordinates defining
\begin{align}
r &=\sqrt{(x^1)^2+(x^2)^2},\nonumber\\
z &=x^3,\nonumber\\
\varphi &=\arctan\frac{x^2}{x^1}. \label{3.26}
\end{align}
We take $e^{i\varphi}$ instead of $\varphi$, for the latter is a multivalued function, and it cannot be promoted to a self-adjoint operator, so that the commutation relations~\eqref{rhoMincommrel} become:
\begin{align}
[x^0,r] &=0,\nonumber\\
[x^0,z] &=0,\nonumber\\
[x^0,e^{i\varphi}] &=\varrho\,\e^{i\varphi}.
\end{align}
In this way, we have two complete sets of commuting observables given by $(r,z,\varphi)$ and $(r,z,x^0)$. On the Hilbert space of $L^2$ functions of the first set, the operators $r,z,\varphi$ act as multiplication operators, while the action of $x^0$ is that of the angular momentum along the 3-axis
\begin{equation}
x^0 \psi(r,z,x^0)=-i \varrho  \partial_\varphi \psi(r,z,x^0).
\end{equation}

Expressing the functions of $\vec x$ in cylindrical coordinates as
\be
\psi(\vec x)=\psi(r ,z,\varphi)=\sum_{n=-\infty}^\infty \psi_n(r,z)\e^{\ii n\varphi} \label{3.31}
\ee
we have that\footnote{Eq.~\eqref{3.28} is valid (convergent) only for a class of functions. However, since this class is sufficiently large, we will ignore this subtlety.}
\be
[x^0,\psi]=\sum_n n \varrho \psi_n(r,z)\e^{\ii n\varphi}=-\ii \varrho\del_\varphi \psi(r,z,\varphi)  = n\varrho \psi \label{3.28}
\ee

Therefore, in this case  the spectrum of time is discrete, being the whole of $\mathbb{Z}$~\cite{lizzivitale}. 
The eigenstates of $\varphi$ are given by a Fourier superposition
\begin{equation}
\delta(\varphi)=\frac{1}{2\pi}\sum^{\infty}_{n=-\infty} \e^{in\varphi}.
\end{equation}

\section{\texorpdfstring{$\varrho$}{}  vs.\ \texorpdfstring{$\kappa$}{}: The quantum groups} \label{rhovskappa}

Central in the discussion on localizability are the symmetry groups of noncommutative spacetimes. In this Section, we will present the quantum group $\varrho$-Poincar\'e\ by means of comparison with the well-known $\kappa$-Poincar\'e\ quantum group, starting by a review of  the latter.

\subsection{The \texorpdfstring{$\kappa$}--Poincar\'e quantum group \texorpdfstring{$\mathcal{C}_\kappa(P)$}{}} \label{kalgebra}
Although historically the $\kappa$-Minkowski spacetime was found starting from  the $\kappa$-Poincar\'e\ Hopf algebra as a quotient  by the Lorentz subgroup, here we will follow the opposite path, i.e., we will find the algebra and the group as symmetries of $\kappa$-Minkowski spacetime. The content of this Subsection is not new; we present it  as a prelude to the $\varrho$ case.

We start by defining  the $\kappa$-Poincar\'e\ $\mathcal{C}_\kappa(P)$\footnote{To be distinguished  from the quantum group $U_\kappa(\mathfrak{p})$ obtained deforming the Hopf algebra of the universal enveloping algebra $\mathfrak{p}$.} as the deformation of the algebra of continuous functions on the Poincar\'e\ group that preserves the $\kappa$-Minkowski commutation relations, i.e., the algebra generated by $\{\Lambda^\mu_\nu, a^\mu \}$ that leaves~\eqref{kappaMincommrel} invariant under the transformation
\begin{equation}
x^\mu \rightarrow x'^\mu ={\Lambda^\mu}_\nu \otimes x^\nu +a^\mu \otimes 1, \label{LorentzTrasf}
\end{equation}
from $\mathcal{M}_\kappa$ to $\mathcal{C}_\kappa(P) \otimes \mathcal{ M}_\kappa$. 
Note that~\eqref{LorentzTrasf} has the form of a left coaction of $\mathcal{C}_\kappa(P)$ on $\mathcal{ M}_\kappa\subset \mathfrak{p}^*$, the latter being the dual of the Poincar\'e\ algebra.
Let us recall that, given an algebra $(\mathcal{A},\mu,\eta)$ and a coalgebra $(\mathcal{C},\Delta,\varepsilon)$,  a left coaction $\beta_L :\mathcal{A} \rightarrow \mathcal{C} \otimes \mathcal{A}$ is a linear mapping satisfying
\begin{subequations}
\begin{align}
(id \otimes \beta_L ) \circ \beta_L &=(\Delta \otimes id) \circ \beta_L & \text{(coassociativity),} \label{1.105a}\\
(\varepsilon \otimes id) \circ \beta_L &=id & \text{(counitality)}. \label{1.105b}
\end{align}
\end{subequations} 
The coaction is said to be covariant if it is an homomorphism:
\begin{subequations}
\begin{align}
\beta_L(ab) &=\beta_L (a) \beta_L (b), \hspace{1cm} a,b\in \mathcal{A},\label{1.106a}\\
\beta_L(1) &=1\otimes 1;\label{1.106b}
\end{align}
\end{subequations}
in this case, it preserves the algebra structure on which it coacts.

We require~\eqref{LorentzTrasf} to be a covariant left coaction. In other words, recalling~\eqref{kappaMincommrel}, and since from~\eqref{1.106a} $\beta_L([x^\mu, x^\nu])=[\beta_L(x^\mu),\beta_L(x^\nu)]$, we ask that
\begin{equation}
[x'^\mu, x'^\nu]=i\lambda ({\delta ^\mu}_0 x'^\nu -{\delta ^\nu}_0 x'^\mu ). \label{2.3}
\end{equation}
By imposing eq.~\eqref{2.3} it is possible to recover part of the full algebra structure of $\mathcal{C}_\kappa(P)$.  Indeed, the left-hand side (LHS) of~\eqref{2.3} yields
\begin{align}
[x'^\mu, x'^\nu] =&[{\Lambda^\mu}_\alpha \otimes x^\alpha +a^\mu \otimes 1, {\Lambda^\nu}_\beta \otimes x^\beta +a^\nu \otimes 1] = \nonumber\\
=&{\Lambda^\mu}_\alpha {\Lambda^\nu}_\beta \otimes x^\alpha x^\beta - {\Lambda^\nu}_\beta {\Lambda^\mu}_\alpha \otimes x^\beta x^\alpha + {\Lambda^\mu}_\alpha a^\nu \otimes x^\alpha - a^\nu {\Lambda^\mu}_\alpha \otimes x^\alpha +\nonumber\\
&+ a^\mu{\Lambda^\nu}_\beta \otimes x^\beta - {\Lambda^\nu}_\beta a^\mu \otimes x^\beta + [a^\mu, a^\nu] \otimes 1. \label{2.4}
\end{align}
while the the right-hand side (RHS) of~\eqref{2.3} assumes the form
\begin{equation}
i\lambda ({\delta ^\mu}_0 x'^\nu -{\delta ^\nu} _0 x'^\mu )=
i\lambda ({\delta ^\mu}_0 ( {\Lambda^\nu}_\sigma \otimes x^\sigma + a^\nu \otimes 1) -{\delta ^\nu} _0 ( {\Lambda^\mu}_\rho \otimes x^\rho + a^\mu \otimes 1) ). \label{2.5}
\end{equation}
Thus, equating terms at order 0 in $x$, it follows straightforwardly
\begin{equation}
[a^\mu, a^\nu]=i\lambda ({\delta ^\mu}_0 a^\nu - {\delta ^\nu} _0 a^\mu), \label{2.6}
\end{equation}
and the translational parameters, unlike the classical Poincar\'e\ group case,  do not commute. This poses problems in localizability of $\kappa$-Poincar\'e\ transformed observables~\cite{lizzimercati}.  These commutation relations are isomorphic to the $\kappa$-Minkowski ones, a feature connected to the bicrossproduct structure of the quantum group~\cite{kmink1}. 

Consider the terms quadratic in $\Lambda$:
\begin{align}
{\Lambda^\mu}_\alpha {\Lambda^\nu}_\beta \otimes x^\alpha x^\beta - {\Lambda^\nu}_\beta {\Lambda^\mu}_\alpha \otimes x^\beta x^\alpha 
=&{\Lambda^\mu}_\alpha {\Lambda^\nu}_\beta \otimes x^\alpha x^\beta - {\Lambda^\nu}_\beta {\Lambda^\mu}_\alpha \otimes x^\beta x^\alpha  \label{2.7}\\
& + {\Lambda^\nu}_\beta {\Lambda^\mu}_\alpha \otimes x^\alpha x^\beta - {\Lambda^\nu}_\beta {\Lambda^\mu}_\alpha \otimes x^\alpha x^\beta\nonumber\\
=&[{\Lambda^\mu}_\alpha , {\Lambda^\nu}_\beta ] \otimes x^\alpha x^\beta + {\Lambda^\nu}_\beta {\Lambda^\mu}_\alpha \otimes i\lambda ({\delta ^\alpha}_0 x^\beta -{\delta ^\beta} _0 x^\alpha ),  \nonumber
\end{align}
from which it  follows, since~\eqref{2.3} has no second order terms in $x$ on the right-hand  side,
\begin{equation}
[{\Lambda^\mu}_\alpha , {\Lambda^\nu}_\beta ] =0. \label{2.8}
\end{equation}
Therefore,  the Lorentz sector remains undeformed,  having trivial commutators. 

We remark that in this discussion we are considering a single particle. One could consider also multiparticle systems described by set of coordinates ${x^\mu}^{(m)}$ and consider commutators $[{x^\mu}^{(m)},{x^\nu}^{(n)}]$. The situation is then more complicated and the coaction is not covariant for ordinary $\kappa$-Minkowski spacetime. Nevertheless, it becomes covariant for a lightlike version of it~\cite{Lizzi:2021rlb}. In this paper, we will remain in the usual one-particle case, but it would be interesting to consider in the $\varrho$-Minkowski setting also the two-particle case.

Let us equate the remaining terms on the left- and right-hand sides:
\begin{equation}
\begin{split}
{\Lambda^\nu}_\beta {\Lambda^\mu}_\alpha \otimes i\lambda ({\delta ^\alpha}_0 x^\beta -{\delta ^\beta} _0 x^\alpha ) +{\Lambda^\mu}_\alpha a^\nu \otimes x^\alpha - a^\nu {\Lambda^\mu}_\alpha \otimes x^\alpha +\\
+ a^\mu{\Lambda^\nu}_\beta \otimes x^\beta - {\Lambda^\nu}_\beta a^\mu \otimes x^\beta
=i\lambda ({\delta ^\mu}_0  {\Lambda^\nu}_\sigma \otimes x^\sigma  -{\delta ^\nu} _0  {\Lambda^\mu}_\rho \otimes x^\rho  ). \label{2.9}
\end{split}
\end{equation}
It is easy to see that this last condition imposes that  the remaining commutators $[\Lambda^\mu_\nu,a^\rho]$ satisfy a compatibility condition:
\begin{equation}
[{\Lambda^\mu}_\alpha,a^\nu]+[a^\mu, {\Lambda^\nu}_\alpha ]=i\lambda\bigl( {\Lambda^\mu}_\alpha({\Lambda^\nu}_0 -{\delta^\nu}_0 ) - {\Lambda^\nu}_\alpha( {\Lambda^\mu}_0-{\delta^\mu}_0 )\bigr)\label{2.10}
\end{equation}
which should be settled by further requests. This is a consequence of the fact that  relations~\eqref{kappaMincommrel} admit more than one single covariance group.

From~\eqref{1.105a} and~\eqref{1.105b}, we can find the coproducts and the counits. Acting with the LHS of~\eqref{1.105a} on $x^\mu$, and recalling~\eqref{1.106b}, we find:
\be
(id \otimes \beta_L)\circ ({\Lambda^\mu}_\nu \otimes x^\nu +a^\mu\otimes 1) ={\Lambda^\mu}_\nu\otimes{\Lambda^\nu}_\alpha \otimes x^\alpha+{\Lambda^\mu}_\nu \otimes a^\nu\otimes 1+a^\mu\otimes 1\otimes 1,
\ee
while from the RHS we have
\be
(\Delta\otimes id)\circ({\Lambda^\mu}_\nu \otimes x^\nu +a^\mu\otimes 1)=\Delta({\Lambda^\mu}_\nu)\otimes x^\nu+\Delta(a^\mu)\otimes 1.
\ee
Comparing the results, we have that
\begin{subequations}
\begin{align}
\Delta (a^\mu)&={\Lambda^\mu}_\nu \otimes a^\nu+ a^\mu \otimes 1, \label{2.32a}\\
\Delta ({\Lambda^\mu}_\nu)&= {\Lambda^\mu}_\alpha \otimes {\Lambda^\alpha}_\nu. \label{2.32b}
\end{align}
\end{subequations}
Turning to~\eqref{1.105b}, and acting on $x^\mu$, we have that
\be
(\varepsilon \otimes id)\circ({\Lambda^\mu}_\nu \otimes x^\nu +a^\mu\otimes 1) =\varepsilon({\Lambda^\mu}_\nu)\otimes x^\nu+\varepsilon(a^\mu)\otimes 1=id(x^\mu)=x^\mu,
\ee
and therefore,
\begin{subequations}
\begin{align}
\varepsilon (a^\mu)&=0, \label{2.32c}\\
\varepsilon ({\Lambda^\mu}_\nu)&={\delta^\mu}_\nu. \label{2.32d}
\end{align}
\end{subequations}
As far the the antipodes are concerned, by the Hopf algebra axioms, it can be shown that they remain undeformed:
\begin{subequations}
\begin{align}
S(a^\mu) &=-a^\nu {(\Lambda^{-1})^\mu}_\nu, \label{2.32e}\\
S({\Lambda^\mu}_\nu) &={(\Lambda^{-1})^\mu}_\nu. \label{2.32f}
\end{align}
\end{subequations}

\subsection{\texorpdfstring{$\mathcal{C}_\kappa(P)$}{}  Structure from the \texorpdfstring{$r$-matrix}{}} \label{rmatrixalgebra}
To fully compute the commutators between coordinate functions of $\mathcal{C}_\kappa(P)$, we may  follow a different  approach   based on the  introduction of the classical $r$-matrix (see Appendix~\ref{App1.4} for details), which will turn to be useful for the $\varrho$-Minkowski case.

A classical $r$-matrix for $\mathcal{C}_\kappa(P)$ is found to be~\cite{koso}
\begin{equation}
r=\ii\lambda M_{0\nu} \wedge P^{\nu} \label{2.11}
\end{equation}
with $M_{\mu\nu}$ and  $P_{\nu}$ the generators of the Poincar\'e\ algebra. 
It can be checked to satisfy the modified Yang-Baxter equation
\begin{equation}
[[r,r]]=\ii \lambda^2 \left (\frac{1}{2} g_{00} M_{\mu \nu} \wedge P^\mu \wedge P^\nu -  M_{\nu 0} \wedge P^\nu \wedge P_0\right), \label{2.13}
\end{equation}
where $[[\cdot, \cdot]]$ denotes the  bracket~\eqref{1.48} described in the Appendix~\ref{App1.4}, while the RHS is invariant under the group action. 
In order to compute the Sklyanin  brackets~\eqref{1.52} of the group parameters,  we need  the Poincar\'e\ left- and right-invariant vector fields. These are obtained starting form the five-dimensional representation\footnote{The arrow indicates four-dimensional vectors.} of  $ISO(1,3)$:
\begin{equation}
g= \begin{pmatrix}
\Lambda & \vec a\\
\vec 0^{\, T} & 1
\end{pmatrix}, \label{2.14}
\end{equation}
through the left- and right-invariant Maurer-Cartan 1-forms 
\bea
\Theta_L &=&g^{-1} d g= \Theta_L^{\alpha\beta} M_{\alpha\beta}+ \Theta_L^\alpha P_\alpha , \label{2.15a}\\
\Theta_R &=&d g g^{-1}=  \Theta_R^{\alpha\beta} M_{\alpha\beta}+ \Theta_R^\alpha P_\alpha. \label{2.15b}
\eea
By duality, the left- and right-invariant vector fields result to be 
\begin{equation}\label{2.25}
\begin{array}{lll}
X_{\alpha \beta}^L &={\Lambda^\mu}_\alpha \frac{\partial}{\partial \Lambda^{\mu\beta}} -{\Lambda^\mu}_\beta \frac{\partial}{\partial \Lambda^{\mu\alpha}},  
& X_\alpha^L ={\Lambda^\mu}_\alpha \frac{\partial}{\partial a^\mu},\\
X_{\alpha \beta}^R &=\Lambda_{\beta\nu} \frac{\partial}{\partial {\Lambda^\alpha}_\nu} -\Lambda_{\alpha\nu} \frac{\partial}{\partial {\Lambda^\beta}_\nu}+a_\beta \frac{\partial}{\partial a^\alpha} -a_\alpha \frac{\partial}{\partial a^\beta},\;\;\;\;\;\;
& X_\alpha^R =\frac{\partial}{\partial a^\alpha},
\end{array}
\ee
which enable us to rewrite~\eqref{1.52} as
\be
\{f,g\} =- \lambda(X_{0\nu}^R\wedge X^{R\nu}-X_{0\nu}^L\wedge X^{L\nu})(\dd f,\dd g). \label{2.27}
\ee
Here we have rescaled the vector fields $X_{\alpha \beta}$ by a factor of $\ii$. 
Performing the calculation for $a^\rho$ and $a^\sigma$:
\be
\{a^\rho,a^\sigma \} =- \lambda \left(a_\nu \frac{\partial}{\partial a^0} -a_0 \frac{\partial}{\partial a^\nu} \right) \wedge \frac{\partial}{\partial a_\nu} (\dd a^\rho,\dd a^\sigma)=-\lambda (a^\sigma {\delta^\rho}_0 -a^\rho {\delta^\sigma}_0). \label{2.28}
\ee
The commutators are then obtained via the canonical prescription $\{,\} \rightarrow \frac{1}{i} [,]$, and we find the previously stated result~\eqref{2.6} quantizing the Poisson-Hopf algebra to a deformed one.

A calculation of $\{{\Lambda^\alpha}_\beta,{\Lambda^\mu}_\nu \}$ gives identically 0, since $P^\mu$ does not contain derivatives in $\Lambda$ neither in left nor in right bases, so the result~\eqref{2.8} comes straightforwardly.

Unlike what we found employing the covariance method, we can now fix the mixed brackets:
\begin{align}
\{{\Lambda^\alpha}_\beta,a^\rho \} =& - \lambda \left( \Lambda_{\nu \mu} \frac{\partial}{\partial {\Lambda^0}_\mu}-\Lambda_{0 \mu} \frac{\partial}{\partial {\Lambda^\nu}_\mu} +a_\nu \frac{\partial}{\partial a^0} -a_0 \frac{\partial}{\partial a^\nu} \right)\wedge \frac{\partial}{\partial a_\nu} (\dd {\Lambda^\alpha}_\beta,\dd a^\rho)\nonumber\\
&+ \lambda \left( {\Lambda^\mu}_0 \frac{\partial}{\partial \Lambda^{\mu \nu}}-{\Lambda^\mu}_\nu \frac{\partial}{\partial \Lambda^{\mu 0}} \right) \wedge \Lambda^{\kappa \nu} \frac{\partial}{\partial a^\kappa} (\dd {\Lambda^\alpha}_\beta,\dd a^\rho)\nonumber\\
=& \lambda (({\Lambda^\alpha}_0 -{\delta^\alpha}_0 ){\Lambda^\rho}_\beta +(\Lambda_{0\beta}-g_{0\beta})g^{\alpha \rho}). \label{2.29}
\end{align}
Considering the commutators,\footnote{Note that the canonical substitution prescription is ordering unambiguous due to the commutativity of the $\Lambda$'s.} we obtain
\begin{equation}
[{\Lambda^\alpha}_\beta,a^\rho]=-\ii \lambda (({\Lambda^\alpha}_0 -{\delta^\alpha}_0 ){\Lambda^\rho}_\beta +(\Lambda_{0\beta}-g_{0\beta})g^{\alpha \rho}). \label{2.30}
\end{equation}
Having completed the algebra structure of $\mathcal{C}_\kappa(P)$, we note that in this formulation the Lorentz sector is undeformed, while the translational one and the cross-relations are noncommutative, giving intuitively an increase in uncertainty of transformed observables.

\subsection{The $\varrho$-Poincar\'e quantum group $\mathcal{C}_\varrho(P)$}\label{rhopoincare}
Following the discussion made in Subsection \ref{rmatrixalgebra}, we will derive the commutation relations (i.e., the algebra sector)  of the $\mathcal{C}_\varrho(P)$ quantum group starting from the classical $r$-matrix of $\varrho$-Minkowski spacetime.

First, note that left- and right-invariant vector fields retain the same expressions~\eqref{2.25}. The only difference with the $\kappa$-Poincar\'e\ quantum group is in the $r$-matrix, which in this case assumes the form~\cite{Lukierski:2005fc, lizzivitale} 

\begin{equation}
r=-i\varrho (P_0 \wedge M_{12}).\label{3.2}
\end{equation}

Note that, unlike the case of the classical $r$-matrix of $\kappa$-Minkowski spacetime which satisfies a modified Yang-Baxter equation (MYBE), \eqref{3.2} satisfies the classical Yang-Baxter equation (CYBE) -- in fact, computing the brackets
\begin{subequations}
\begin{align}
[r_{12},r_{13}] &=-\varrho^2[M_{12},M_{12}]\otimes P_0\otimes P_0=0,\nonumber\\
[r_{12},r_{23}] &=\varrho^2 P_0\otimes[M_{12},M_{12}]\otimes P_0=0,\nonumber\\
[r_{13},r_{23}] &=-\varrho^2 P_0\otimes P_0\otimes [M_{12},M_{12}]=0,\label{3.3}
\end{align}
\end{subequations}
and thus, $[[r,r]]=0$.

The Sklyanin bracket~\eqref{1.52} assumes the form
\begin{equation}
\{f,g\} =-\varrho(X_{12}^R\wedge X_0^{R}-X_{12}^L\wedge X_0^{L})(df,dg),\label{3.4}
\end{equation}
so that we can compute the brackets between Poincar\'e\ coordinates as done earlier:
\begin{align}
\{ \alpha^\mu, \alpha^\nu \} & =-\varrho [{\delta^\nu}_0 (a_2 {\delta^\mu}_1-a_1 {\delta^\mu}_2) -{\delta^\mu}_0 (a_2 {\delta^\nu}_1-a_1 {\delta^\nu}_2)],\nonumber \\
\{ {\Lambda^\mu}_\nu, {\Lambda^\varrho}_\sigma \} &=0,\nonumber\\
\{ {\Lambda^\mu}_\nu, a^\varrho \} &=-\varrho \left[{\delta^\varrho}_0 (\Lambda_{2\nu} {\delta^\mu}_1 -\Lambda_{1\nu} {\delta^\mu}_2) -{\Lambda^\varrho}_0 ({\Lambda^\mu}_1 g_{2\nu} -{\Lambda^\mu}_2 g_{1\nu} )\right].\label{3.5}
\end{align}
Therefore, the commutators are
\begin{subequations}
\begin{align}
\left[ a^\mu, a^\nu \right] &=-i \varrho [{\delta^\nu}_0 (a_2 {\delta^\mu}_1-a_1 {\delta^\mu}_2) -{\delta^\mu}_0 (a_2 {\delta^\nu}_1-a_1 {\delta^\nu}_2)],\label{3.6a}\\
\left[ {\Lambda^\mu}_\nu, {\Lambda^\varrho}_\sigma \right] &=0,\label{3.6b}\\
\left[ {\Lambda^\mu}_\nu, a^\varrho \right] &=-i \varrho \left[{\delta^\varrho}_0 (\Lambda_{2\nu} {\delta^\mu}_1 -\Lambda_{1\nu} {\delta^\mu}_2) -{\Lambda^\varrho}_0 ({\Lambda^\mu}_1 g_{2\nu} -{\Lambda^\mu}_2 g_{1\nu} )\right].\label{3.6c}
\end{align}
\end{subequations}
Again it is easy to see that the commutation relations between $a^\mu$ and $a^\nu$ reproduce Eqs.~\eqref{rhoMincommrel}, and $\varrho$-Minkowski spacetime can therefore be recovered from the momenta sector of $\mathcal{C}_\varrho(P)$. Moreover, it can be checked that the commutation relations of the $\varrho$-Minkowski spacetime~\eqref{rhoMincommrel} are covariant under the left coaction~\eqref{LorentzTrasf} if the commutation relations~\eqref{3.6a}-\eqref{3.6c} are implemented. 

For the coalgebra sector and the antipode, since the left coaction is the same as that of the $\kappa$-Poincar\'e\ quantum group, they retain the forms~\eqref{2.32a},~\eqref{2.32b}, \eqref{2.32c}, \eqref{2.32d}, \eqref{2.32e}, and \eqref{2.32f}. It is then trivial to see that taking the limit $\varrho \rightarrow 0$, the classical commutative case is recovered.

The fundamental result of this analysis is that the algebra sector of the translational parameters and the cross--relations between translational and Lorentz parameters are noncommutative~\cite{lizzivitale}. This will lead to an increase in uncertainty in $\varrho$-Poincar\'e\ transformations, as we will show in the following.

\section{Localizability in \texorpdfstring{$\varrho$}{}-Minkowski space} \label{section3}
We now analyse localizability in the $\varrho$-Minkowski space, following what has been done in~\cite{lizzimercati} for the $\kappa$ case. We first consider coordinate localizability features coming from~\eqref{3.24}, then we realize the elements of the quantum group on a suitable Hilbert space, we derive uncertainty relations for them, and we discuss localizability in $\varrho$-Minkowski in relation to observers and observables.

\subsection{Localized states in $\mathcal{M}_\varrho$}

Let us suppose to sharply measure an eigenvalue $\varrho \bar{n}$ of the time operator. The system would be in an eigenstate of time $\bar{\chi}(\varphi)=e^{i\bar{n}\varphi}$, so that we would have complete delocalization in $\varphi$. If the measure has instead some degree of uncertainty in time, we would have a finite sum over the available elements of the basis, and this would give, in turn, a degree of uncertainty in $\varphi$, as in the ordinary quantum mechanical angular momentum theory.

From~\eqref{3.24}, we expect, however, that sharp spacetime localization is possible in the case $\langle x^1\rangle=\langle x^2\rangle=0$. In our cylindrical coordinates, this corresponds to perfect localization in $\langle r\rangle =0$. Since $r$ commutes with $z$ and $x^0$  we can find a state that  localizes in $r$ as well as in $z$ and $x^0$. As usual, the state will not be a proper square integrable vector, but a $\delta$-like distribution reachable via a limiting process. A state of this kind can be constructed as
\begin{equation}
\psi_{n_0}(r,z,x^0)=\frac{1}{2\pi} \int\limits^\pi_{-\pi} \dd\varphi  \,\e^{-i(n-n_0)\varphi}\;\xi(r,z) ,
\end{equation}
where the integral yields  a $\delta(n-n_0)$ that gives a state localized in time at $n_0$, and $\xi(r,z)$ is a function of $r$ and $z$ localized around  $(r_0 ,z_0)$. This can be taken to be  a factorized product of two states in the Hilbert space (e.g., Gaussian distributions) that tend to delta distributions in the limit of their amplitudes going to  $0$ (e.g., the Gaussian variances $\rightarrow 0$).

From~\eqref{3.26}, $x^1=r\cos\varphi$, $x^2=r\sin \varphi$, but $\varphi$ is completely undetermined since we are in an eigenstate of $x^0$. Computing the mean values on the state, we have $\langle x^1 \rangle= r_0\cos\varphi$ and $\langle x^2\rangle =r_0 \sin\varphi$; hence, perfect localization in $x^\mu$ is possible only if $r_0=0$. We obtain then a 2-parameter localized family of states $|o_{n,z}\rangle$.
In the particular case of $n_0=z_0=0$ we can define a localized origin state $|o\rangle$.
This result is analogous with the case of $\kappa$-Minkowski spacetime~\cite{lizzimercati, lizzimercati2}, for which it was found that a one-parameter family of localized states $|o_\tau \rangle$ does exist, allowing for the definition of a localized origin state $|o\rangle$.

Let us note here an  important fact. The following function also gives a localized state at time $n_0+\alpha$
\begin{equation}
\psi_{n_0+\alpha}(r,z,x^0)=\frac{1}{2\pi}\int\limits^\pi_{-\pi} \dd\,\varphi \e^{-i(n-n_0+\alpha)\varphi}\;\xi(r,z)
\end{equation}
which is only  periodic in $\varphi$ {up to the phase $\e^{i2\pi\alpha}$}. This means that these two states belong to different domains of self-adjointness of the operator $x^0$. This aspect will be discussed elsewhere.

\subsection{$\varrho$-Poincar\'e\ realization}

Since later on we will deal with localization properties of the quantum group parameters, we now present a realization for the $\varrho$-Poincar\'e\ group, following the approach carried on for the $\kappa$-Poincar\'e\ group in~\cite{lizzimercati}. 

We start noting that, as in $\kappa$-Poincar\'e\ case, the $\Lambda$'s commute with each other, and so they can be realized   classically.
In terms of  the infinitesimal generators of the Lorentz group ${\omega^{\mu}}_\nu$, we have that
\begin{equation}
{\Lambda^\mu}_\nu={(\exp \omega)^\mu}_\nu, \label{2.70}
\end{equation}
with the auxiliary antisymmetry condition
\; $
{\omega^\mu}_\varrho g^{\varrho \nu}=-{\omega^\nu}_\varrho g^{\varrho \mu}.
$ 

For the $a$'s, by considering the commutation relation~\eqref{3.6c}, we formulate the ansatz
\begin{equation}
a^\varrho=i \varrho \left[{\delta^\varrho}_0 (\Lambda_{2\nu} {\delta^\mu}_1 -\Lambda_{1\nu} {\delta^\mu}_2) -{\Lambda^\varrho}_0 ({\Lambda^\mu}_1 g_{2\nu} -{\Lambda^\mu}_2 g_{1\nu} )\right]\frac{\partial}{\partial {\Lambda^\mu}_\nu}. \label{3.36}
\end{equation}
To have a realization of the group,  we must show that this is coherent with~\eqref{3.6a}.
From~\eqref{3.6a}, it has to be 
\begin{equation}
\begin{split}
\left[ a^\mu, a^\nu \right] =&-\varrho^2(-{\delta^\mu}_0\Lambda_{20}{\delta^\nu}_1+{\delta^\mu}_0\Lambda_{10}{\delta^\nu}_2+{\delta^\nu}_0\Lambda_{20}{\delta^\mu}_1-{\delta^\nu}_0\Lambda_{10}{\delta^\mu}_2)\times\\
&\times({\Lambda^\alpha}_1g_{2\beta}-{\Lambda^\alpha}_2g_{1\beta})\frac{\partial}{\partial {\Lambda^\alpha}_\beta}.\label{3.38}
\end{split}
\end{equation}
On computing the LHS, we find 
\begin{equation}
\begin{split}
[a^\varrho,a^\sigma]=\varrho^2 & \left[ {\delta^\varrho}_0(\Lambda_{20}{\delta^\sigma}_1-\Lambda_{10}{\delta^\sigma}_2)({\Lambda^\delta}_2 g_{1\lambda}-{\Lambda^\delta}_1 g_{2\lambda})+\right.\\
&-\left.{\delta^\sigma}_0(\Lambda_{20}{\delta^\varrho}_1-\Lambda_{10}{\delta^\varrho}_2)({\Lambda^\delta}_2 g_{1\lambda}-{\Lambda^\delta}_1 g_{2\lambda})\right]\frac{\partial}{\partial {\Lambda^\delta}_\lambda} \label{3.40}
\end{split}
\end{equation}
which is in agreement with Eq.~\eqref{3.38};  therefore Eqs.~\eqref{2.70} and~\eqref{3.36} give a true realization of $\varrho$-Poincar\'e quantum group.

Finally, in analogy with the $\kappa$-case (cfr.~\cite{lizzimercati}), we add to Eq.~\eqref{3.36} the realization of $\varrho$-Minkowski Eq.~\eqref{3.25}:
\begin{equation}
\begin{split}
a^\varrho&=i \frac{\varrho}{2} \left[{\delta^\varrho}_0 (\Lambda_{2\nu} {\delta^\mu}_1 -\Lambda_{1\nu} {\delta^\mu}_2) -{\Lambda^\varrho}_0 ({\Lambda^\mu}_1 g_{2\nu} -{\Lambda^\mu}_2 g_{1\nu} )\right]\frac{\partial}{\partial {\Lambda^\mu}_\nu}+\\
&+i\frac{\varrho}{2}[{\delta^\varrho}_iq^i-{\delta^\varrho}_0(q^1\partial_2-q^2\partial_1)]+h.c.\label{3.41}
\end{split}
\end{equation}
defined on the Hilbert space $L^2(SO(1,3)\times \mathbb{R}^3)$.

\subsection{$\varrho$-Poincar\'e\ parameters, localization and constraints on transformations}\label{paramloc}
Since the symmetry group of $\varrho$-Minkowski spacetime is deformed according to Eqs.~\eqref{3.6a}-\eqref{3.6c}, we expect localization problems to arise also in observer transformations. Indeed, 
we obtain uncertainty relations in the form
\begin{subequations}\begin{align}
\Delta a^\mu \Delta a^\nu &\geq  \frac{\varrho}{2} |{\delta ^\nu}_0 ( \langle a_2 \rangle  {\delta^\mu}_1 -\langle a_1\rangle {\delta^\mu}_2) - {\delta ^\mu} _0 ( \langle a_2 \rangle {\delta^\nu}_1-\langle a_1 \rangle {\delta^\nu}_2 )|,\label{3.35a}\\
\Delta {\Lambda^\mu}_\alpha \Delta {\Lambda^\nu}_\beta  &\geq  0,\label{3.35b}\\
\Delta {\Lambda^\mu}_\nu \Delta a^\rho &\geq  \frac{\varrho}{2} |{\delta^\varrho}_0 ( \langle \Lambda_{2\nu} \rangle {\delta^\mu}_1 -\langle \Lambda_{1\nu} \rangle {\delta^\mu}_2) -\langle {\Lambda^\varrho}_0 {\Lambda^\mu}_1\rangle g_{2\nu} +\langle{\Lambda^\varrho}_0{\Lambda^\mu}_2\rangle g_{1\nu} |. \label{3.35c}
\end{align}
\end{subequations}
Let us analyze the localization properties of this algebra structure.
We start with the case of pure $\varrho$-Lorentz transformations, i.e., transformations for which translational parameters are sharply localized in 0 ($\langle a^\mu \rangle=0$, $\Delta a^\mu=0$). The relevant constraint on localizability comes from~\eqref{3.35c}:
\be
{\delta^\varrho}_0 ( \langle \Lambda_{2\nu} \rangle {\delta^\mu}_1 -\langle \Lambda_{1\nu} \rangle {\delta^\mu}_2) -\langle {\Lambda^\varrho}_0 {\Lambda^\mu}_1\rangle g_{2\nu} +\langle{\Lambda^\varrho}_0{\Lambda^\mu}_2\rangle g_{1\nu}=0.\label{lorentzloc}
\ee
This, like the case of $\kappa$-Poincar\'e quantum group~\cite{mercati}, admits a solution\footnote{We consider a state $|\phi\rangle$ on wich $\Delta a^\mu=0$. Then, if we take an eigenstate $|\phi_\lambda\rangle$ of ${\Lambda^\mu}_\nu$, we have that, since the $\Lambda$'s commute, their eigenvalues on $|\phi_\lambda\rangle$ are classical Lorentz parameters ${\lambda^\mu}_\nu$. It is possible to show that the only solution of~\eqref{lorentzloc} is ${\lambda^\mu}_0={\delta^\mu}_0$, ${\lambda^1}_3={\lambda^2}_3=0$, ${\lambda^1}_1={\lambda^2}_2$ for every eigenstate such that $\langle \phi|\phi_\lambda \rangle \neq 0$.} for $\langle {\Lambda^\varrho}_0 \rangle ={\delta^\varrho}_0$, $\langle {\Lambda^3}_1 \rangle =\langle {\Lambda^3}_2\rangle=0$, $\langle {\Lambda^1}_1 \rangle =\langle {\Lambda^2}_2\rangle $, and so the only admitted pure $\varrho$-Lorentz transformations are rotations around the 3-axis and the identical tranformation, and they can be sharply localized. For the $\kappa$-Poincar\'e\ quantum group, a slightly different result was found in~\cite{mercati}, namely, that just pure boosts are not admitted,  in accord with~\cite{noboost}. 

For the case of pure translations, i.e., $\langle {\Lambda^\mu}_\nu \rangle={\delta^\mu}_\nu$ and $\Delta{\Lambda^\mu}_\nu=0$, substituting in~\eqref{3.35c} we see that the relation is automatically satisfied, and the only relevant condition is~\eqref{3.35a}. Since $a^3$ is central in the algebra, pure translations along the 3-axis do exist without issues and can be sharply localized.

Considering a pure time translation, the conditions to impose on~\eqref{3.35a} are that $\langle a^i\rangle=0$ and $\Delta a^i=0$, and the equation is trivially satisfied, meaning that pure time translations do exist and can be localized. For pure translations along the 1- and 2-axes the result is different: if we consider, for example, the first case, one would have $\langle a^2 \rangle=0$ that is compatible with $\Delta x^0=0$, but this last condition imposes also that $\langle a^1 \rangle=0$, the same being true switching $a^1$ and $a^2$. This means that the $\varrho$-Poincar\'e\ quantum group admits only pure time translations and pure space translations along the 3-axis.
For comparison, in the $\kappa$ case, it was found that the only possible pure translation is the temporal one.

Summarizing the localization features of the quantum group, the only transformations that can be sharply localized are translations along $x^0$, translations along $x^3$, rotations around $x^3$, and their combinations.

As a special case, we turn our attention to the identical transformation $\langle a^\mu \rangle =0$, $\langle {\Lambda^\mu}_\nu \rangle ={\delta^\mu}_\nu$, $\Delta a^\mu=0$, $\Delta{\Lambda^\mu}_\nu=0$; as we expect the uncertainty relations are satisfied, and therefore, the identity in the $\varrho$-Poincar\'e\ quantum group is a well-defined sharp state.

\subsection{Observers, observables and uncertainties on $\varrho$-Poincar\'e}
Let us  analyze the uncertainties in Poincar\'e\ transformations~\eqref{LorentzTrasf} coming from the deformation features of the quantum group. Since our transformation is a left coaction from $x \in \mathcal{M}_\varrho $ to $x' \in \mathcal{C}_\varrho(P) \otimes \mathcal{M}_\varrho$, we want to find a realization of the tensor product $\mathcal{C}_\varrho(P) \otimes \mathcal{M}_\varrho$. It is convenient to lift $x\in \mathcal{M}_\varrho$ to $1\otimes x\in \mathcal{C}_\varrho(P)\otimes \mathcal{M}_\varrho$.
We can find the action of elements $x'^\mu \in \mathcal{C}_\varrho(P)\otimes \mathcal{M}_\varrho$ on functions $f(\omega,q,x)\in L^2(SO(1,3)\times \mathbb{R}^3_q)\times L^2(\mathbb{R}^3_x)\sim L^2(SO(1,3)\times \mathbb{R}^3_q \times \mathbb{R}^3_x)$ by means of the direct sum of realizations~\eqref{3.25} and~\eqref{3.41}:
\begin{align}
x'^\varrho f(\omega,q,x)=&i\varrho {\Lambda^\varrho}_\sigma [{\delta^\sigma}_ix^i-{\delta^\sigma}_0(x^1\partial_{x_2}-x^2\partial_{x_1})]f(\omega,q,x)+\nonumber\\
&+i \frac{\varrho}{2} \left[{\delta^\varrho}_0 (\Lambda_{2\nu} {\delta^\mu}_1 -\Lambda_{1\nu} {\delta^\mu}_2) -{\Lambda^\varrho}_0 ({\Lambda^\mu}_1 g_{2\nu} -{\Lambda^\mu}_2 g_{1\nu} )\right]\frac{\partial}{\partial {\Lambda^\mu}_\nu}f(\omega,q,x)+\nonumber\\
&+i\frac{\varrho}{2}[{\delta^\varrho}_iq^i-{\delta^\varrho}_0(q^1\partial_{q_2}-q^2\partial_{q_1})]f(\omega,q,x)+\frac{1}{2}h.c.
\end{align}
The Hilbert space admits separable states of the kind
\begin{equation}
|\phi,\psi\rangle =|\phi \rangle \otimes |\psi \rangle,
\end{equation}
with $|\phi \rangle \in L^2(SO(1,3)\times \mathbb{R}_q^3)$ and $|\psi \rangle \in L^2(\mathbb{R}_x^3)$ normalized according to $\langle \phi |\phi \rangle=1$, $\langle \psi |\psi \rangle=1$.

We are ready to give an interpretation of the realization constructed here. We define $L^2(SO(1,3)\times \mathbb{R}_q^3)$ as the space of states of an observer (i.e., the space of $\varrho$-Poincar\'e\ states) and $L^2(\mathbb{R}_x^3)$ as the space of observables (i.e., the space of states of $\varrho$-Minkowski spacetime); furthermore, we assume that a generic state can be realized as a separable element $|\phi,\psi\rangle =|\phi \rangle \otimes |\psi \rangle$, a reasonable assumption since it reflects the fact that the relation between two inertial observers does not depend on the observed state.

The point here is that  we  have at the same time  a noncommutative spacetime on which observables are defined and a noncommutative observer state space, meaning that in general a $\varrho$-Poincar\'e\ transformation between different observers could decrease localizability of states.

Taking into account~\eqref{LorentzTrasf}, and interpreting $x^\mu$ as the coordinates of an inertial observer $\mathcal{O}$, while $x'^{\mu}$ as those of a transformed observer $\mathcal{O'}$, the mean value of the coordinates of a transformed observer would be
\begin{equation}
\langle x'^\mu \rangle=\langle \phi |\otimes \langle \psi | ({\Lambda^\mu}_\nu \otimes x^\nu +a^\mu \otimes 1)|\phi \rangle \otimes |\psi \rangle = \langle \phi |{\Lambda^\mu}_\nu |\phi \rangle \langle \psi | x^\nu |\psi \rangle +\langle \phi | a^\mu |\phi \rangle,
\end{equation}
while for the uncertainties of transformed states in relation to those of the starting ones, we can write
\begin{equation}
\Delta(x'^\mu)^2=\langle (x'^\mu)^2\rangle -\langle x'^\mu \rangle^2=\Delta ({\Lambda^\mu}_\nu \otimes x^\nu)^2+\Delta(a^\mu)^2+2cov({\Lambda^\mu}_\nu,a^\mu)\langle x^\nu \rangle,
\end{equation}
since $\langle a\otimes b \rangle=\langle a \rangle \otimes \langle b \rangle$ and the covariance between elements on different sides of the tensor product is 0.
In the following we will specialise the construction to three notable cases, that is, identity transformations, origin states transformations, and translations.

\subsubsection{Identity transformation state}
Since we know from our analysis that a sharp identity state does exist in the $\varrho$-Poincar\'e quantum group, we can consider identity transformations.
We define the identity state $|i\rangle$ for our realization of $\mathcal{C}_\varrho(P)$ as follows:
\begin{equation}
\langle i|f(a,\Lambda)|i\rangle=\varepsilon(f)
\end{equation}
with   $f(a,\Lambda)\in \mathcal{C}_\varrho(P)$.
Then the   state
\begin{equation}
|i, \psi\rangle=|i \rangle \otimes |\psi \rangle
\end{equation}
can be linked to the $\varrho$-Poincar\'e\ transformation between two coincident observers, as one can see working the following calculation:
\begin{equation}
\langle x'^\mu \rangle=\langle i |\otimes \langle \psi | ({\Lambda^\mu}_\nu \otimes x^\nu +a^\mu \otimes 1)|i \rangle \otimes |\psi \rangle = \langle i |{\Lambda^\mu}_\nu |i \rangle \langle \psi | x^\nu |\psi \rangle +\langle i | a^\mu |i \rangle;
\end{equation}
but recalling the counits~\eqref{2.32c} and~\eqref{2.32d}
\begin{equation}
\langle x'^\mu \rangle=  \langle \psi | x^\mu |\psi \rangle.
\end{equation}
The same result is achieved for a generic monomial in coordinates $x'^{\mu_1}\cdots x'^{\mu_n}$:
\begin{equation}
\begin{split}
\langle x'^{\mu_1}\cdots x'^{\mu_n} \rangle =&\langle i |\otimes \langle \psi | x'^{\mu_1}\cdots x'^{\mu_n}|i \rangle \otimes |\psi \rangle=\\
=&\langle i|a^{\mu_1} \cdots a^{\mu_n} |i\rangle +\langle i|\mathcal{O}^{\mu_1 \cdots \mu_n}_\nu(a,\Lambda)|i\rangle \langle \psi  |x^\nu|\psi\rangle +\\
&+\cdots +\langle i|\mathcal{O}^{\mu_1 \cdots \mu_n}_{\nu_1 \cdots \nu_n} (a,\Lambda)|i\rangle \langle \psi  |x^{\nu_1}\cdots x^{\nu_n}|\psi\rangle,
\end{split}
\end{equation}
with $\mathcal{O}(a,\Lambda)$ generic monomials in $a$'s and $\Lambda$'s. Since the counit map is a homomorphism, every monomial that contains at least one $a$ vanishes ($\varepsilon(a^\mu)=0$), and the only surviving term is the one  with an equal number of upper and lower indices that is a product of $\Lambda$'s only. Again from the homomorphism property, one obtains that $\varepsilon(\mathcal{O}^{\mu_1 \cdots \mu_n}_{\nu_1 \cdots \nu_n} (a,\Lambda))={\delta^{\mu_1}}_{\nu_1} \cdots {\delta^{\mu_n}}_{\nu_n}$, and
\begin{equation}
\langle x'^{\mu_1}\cdots x'^{\mu_n} \rangle=\langle \psi | x^{\mu_1}\cdots x^{\mu_n} |\psi \rangle.
\end{equation}
Then  one easily sees that uncertainties between the two events coincide:
\begin{equation}
\Delta(x'^\mu)^2=\langle (x'^\mu)^2\rangle -\langle x'^\mu \rangle^2=\langle ({\delta^\mu}_\nu x^\nu)^2\rangle -\langle x^\mu \rangle^2= \Delta(x^\mu)^2.
\end{equation}
Coincident observers are well defined in $\varrho$-Minkowski spacetime and they agree on every measurement they make. These results are identical to those found in~\cite{lizzimercati} for $\kappa$-Minkowski spacetime.

\subsubsection{Origin state transformations}\label{oritrans}
We ask what  the   observer $\mathcal{O}'$  measures  after $\varrho$-Poincar\'e\ transforming the origin state; the starting state is
\begin{equation}
|\phi, o\rangle=|\phi \rangle \otimes |o \rangle,
\end{equation}
therefore,
\begin{equation}
\langle x'^\mu \rangle=\langle \phi |\otimes \langle o | ({\Lambda^\mu}_\nu \otimes x^\nu +a^\mu \otimes 1)|\phi \rangle \otimes |o \rangle = \langle \phi |{\Lambda^\mu}_\nu |\phi \rangle \langle o | x^\nu |o \rangle +\langle \phi | a^\mu |\phi \rangle.
\end{equation}
Recalling that $\langle o | x^\mu |o \rangle =0$, we have 
\begin{equation}
\langle x'^\mu \rangle= \langle \phi | a^\mu |\phi \rangle.
\end{equation}
This result entails the fact that the two observers $\mathcal{O}$ and  $\mathcal{O}'$ are comparing positions and not directions, so the expectation value is determined only by the mean value of translation operators.

It can be shown by an analogous computation that the result remains true also for a generic monomial in coordinates $x'^{\mu_1}\cdots x'^{\mu_n}$; in fact, since $\langle o|x^{\mu_1}\cdots x^{\mu_n} |o\rangle=0$ $\forall n$,
\begin{equation}
\langle x'^{\mu_1}\cdots x'^{\mu_n} \rangle=\langle \phi |\otimes \langle o | x'^{\mu_1}\cdots x'^{\mu_n}|\phi \rangle \otimes |o \rangle =\langle \phi | a^{\mu_1}\cdots a^{\mu_n} |\phi \rangle.
\end{equation}
In this case, the uncertainty of the transformed event coincides with that of the translation operator:
\begin{equation}
\Delta(x'^\mu)^2=\langle (x'^\mu)^2\rangle -\langle x'^\mu \rangle^2=\langle (a^\mu)^2\rangle -\langle a^\mu \rangle^2=\Delta(a^\mu)^2.
\end{equation}
Comparing with the  $\kappa$-case~\cite{lizzimercati}, we notice that when  the translational parameter can be localized, in both cases the uncertainty on the final state is zero.   For the $\varrho$-Poincar\'e\ quantum group, recalling~\eqref{3.35a}, this occurs when $\langle a^1\rangle =\langle a^2 \rangle=0$, namely,  for pure translations along $a^0, a^3$ or even mixed translations in $a^0$, $a^3$, while for  the $\kappa$-Poincar\'e\ quantum group, this occurs only for  pure temporal translations.
\subsubsection{Translations}
Another interesting case is that of a pure translation $x'^\mu=1\otimes x^\mu + a^\mu \otimes 1$ of a generic state.
To demonstrate that states $|\phi_T\rangle$ corresponding to  translations do exist in $L^2(SO(1,3)\times \mathbb{R}^3_q)$,  it is necessary to take a sequence of functions which converge to a $\delta$ for the diagonal elements of $\Lambda$ and to 0 for off-diagonal ones, so that $\langle \phi_T|{\Lambda^\mu}_\nu| \phi_T\rangle = {\delta^\mu}_\nu$. We observe that  taking such states and (co)acting with the usual coaction~\eqref{LorentzTrasf}, it is the same thing as (co)acting  on a generic state of  $L^2(SO(1,3)\times \mathbb{R}^3_q)\times L^2(\mathbb{R}^3_x)$ with $x'^\mu=1\otimes x^\mu + a^\mu \otimes 1$.
The expectation value is then 
\begin{equation}
\langle x'^\mu \rangle=\langle \phi |\otimes \langle \psi | (1\otimes x^\mu + a^\mu \otimes 1)|\phi \rangle \otimes |\psi \rangle = \langle \psi | x^\mu |\psi \rangle +\langle \phi | a^\mu |\phi \rangle,
\end{equation}
while the variance
\begin{equation}
\begin{split}
\Delta(x'^\mu)^2 &=\langle (x^\mu)^2+(a^\mu)^2+x^\mu a^\mu +a^\mu x^\mu \rangle -\langle x^\mu \rangle^2 -\langle a^\mu \rangle^2 -2\langle x^\mu \rangle \langle a^\mu \rangle=\\
& =\Delta(x^\mu)^2+\Delta(a^\mu)^2 \geq \Delta(x^\mu)^2.
\end{split}
\end{equation}
Therefore, one sees that acting with a pure translation leads, in general, to  an increase in the state uncertainty.  As for the comparison with the $\kappa$ case, the same considerations apply as those at the end of Subsection \ref{oritrans}.

\section{Conclusions and Outlook} \label{conc}
We have analyzed the localization features of spacetime states of $\mathcal{M}_\varrho$ as well as those of the quantum group $\mathcal{C}_\varrho(P)$ and their consequences on Poincar\'e-deformed transformations.

The main difference between the $\kappa$- and $\varrho$-Minkowski spacetimes is in the nature of the
commutation relations. While for the former these are clearly of radial nature, for the latter 
they are explicitly of an angular one. In the first case, there are no central cartesian coordinates,
while in the second case, $x^3$ commutes with every other one. It is therefore legitimate to think
that this coordinate can be determined without any uncertainty and will not pose problems for its localizability. We have shown that perfect localization of observable states can be achieved in the ``special position'' $x^1 = x^2 = 0$, in accord with the angular nature of the only nonmultiplicative operator $x^0$ that acts as an angular momentum along the 3-axis.

Turning our attention to the issue of symmetries, we have shown that the deformed nature of the Poincar\'e\ quantum groups leads to the interesting feature of having uncertainties arising from deformed Poincar\'e\ transformations. This implies that two different observers will, in general, not agree on the localizability properties of the same state. The localizability properties of the quantum groups can be analyzed by writing uncertainty relations between the noncommutative group parameters. These relations, surprisingly, pose constraints on the admissible deformed-Poincar\'e\ transformations; for example, we have seen that for $\varrho$-Poincar\'e\ pure space translations along the 1- and 2-axes and pure Lorentz transformations are not allowed except for the rotations around the 3-axis. These features were previously discussed for the $\kappa$-case, leading to the so-called ``no-pure'' features of the quantum group~\cite{mercati, noboost}.

It is worth noticing  that in~\cite{lizzimercati}, a particular mixed transformation in the (1+1)-dimensional $\kappa$ case,  leading to a decrease in uncertainty was found. It would be interesting to see if also the $\varrho$-Poincar\'e\ quantum group admits some transformation of this kind and to give some physical interpretation to it.

We have not considered the dual picture of the quantum universal enveloping algebra $U_\varrho (\mathfrak{p})$  that was obtained in~\cite{{Ciric:2017rnf}, DimitrijevicCiric:2018blz} within the twist approach. In that framework, the Lie algebra sector is naturally  underformed, whereas the cosector is modified. There is, however, the possibility of finding a non-linear change of basis which could lead to a different quantum universal enveloping algebra $\widetilde U_\varrho (\mathfrak{p})$ with a bicrossproduct structure,  in analogy with the $\kappa$-Poincar\'e\  case. This could have interesting physical implications  (such as consequences on deformed infinitesimal symmetries and deformed dispersion relations) and it is presently under investigation. 
\appendix

\section{States, observables and observers} \label{AppA}
We review in this Appendix the notions of states, observables and observers, which hold true both in  the commutative and the noncommutative cases.
\paragraph{States.}
A \textit{state} $\phi$ is a linear functional from a $C^*$-algebra $\mathcal{C}$ to the complex field (see, for example,~\cite{WSS}):
\begin{equation}
\phi:\mathcal{C}\rightarrow \mathbb{C},\label{1.112}
\end{equation}
positive defined
\begin{equation}
\phi (a^*a)\geq 0, \hspace{0.5cm} \forall a\in \mathcal{C},\label{1.113}
\end{equation}
and normalized
\begin{equation}
\parallel \phi \parallel=\sup\limits_{\parallel a\parallel \leq 1} \{\phi(a)\}=1.\label{1.114}
\end{equation}
The space of states can be shown to be convex. Any state that can be expressed as a convex combination is said to be a \textit{mixed state}, while states that cannot are called \textit{pure states}.

From a commutative algebra and its set of pure states, it is possible to define a topology and thus obtain the associated topological space through the so-called Connes construction (see~\cite{WSS} for details). Furthermore, we can associate the notion of (functional) states to that of \textit{vector states} on a Hilbert space via a  Gelfand-Naimark-Segal (GNS) construction. Given, in fact, an algebra of bounded operators $\mathcal{B(H)}$ on a Hilbert space $\mathcal{H}$, any normalized vector $|\xi\rangle$ defines a state with expectation value $\phi_\xi(a)=\langle\xi|\hat{a}|\xi\rangle$, $\hat{a}\in\mathcal{B(H)}$. On the contrary, to any state $\phi$ it corresponds a vector state $\xi_\phi\in \mathcal{H}$ such that $\langle\xi_\phi|\hat{a}|\xi_\phi\rangle=\phi(a)$. If the variance $\Delta(a)=\sqrt{\phi(a^2)-\phi(a)^2}=\sqrt{\langle\xi_\phi|\hat{a}^2|\xi_\phi\rangle-(\langle\xi_\phi|\hat{a}|\xi_\phi\rangle)^2}$ is equal to zero, the state is said to be \textit{localized}.

\paragraph{Observables.}
An \textit{observable} $\mathcal{A}$ is, heuristically, a physical quantity that can be measured. Formally, in classical mechanics it is defined as a real-valued function on the phase space, while in quantum mechanics as a self-adjoint operator defined on a Hilbert space.
Therefore, in the present context,  an observable $\mathcal{A}$ is a self-adjoint element of the $C^*$-algebra $\mathcal{C}$. In this way, we can say that a state is a mapping from physical observables to their measured value.

\paragraph{Observers.}
The notion of observer is a more subtle one. Loosely speaking, an observer in classical mechanics is something that performs a measure on a physical system and associates a real numerical value to the corresponding observable function; in quantum mechanics instead, it is a filter procedure that sends, after having performed a measure on a quantum object, a quantum state to a classical one associating numerical eigenvalues to observable operators with discrete spectra, or continuous density eigenvalues to operators with continuous spectra. In  this work, we avoid the problem of giving a rigorous definition by relating  an observer to its reference frame.

An \textit{observer} $\mathcal{O}$ is a reference frame with respect to which the ordinary theory of measurement (i.e., the possibility of finding mean values, variances and other higher moments of one or more observables in a state) can be applied. 
As a final remark, let us notice that,  since we are dealing with special-relativistic theories, not taking into account general relativity (GR) features, we always mean  inertial observers.

\section{Deformation of Hopf algebras} \label{App1.4}
In this Appendix, we recall basic facts about  the classical $r$-matrix~\cite{charipressley}, as well as an approach~\cite{ZakrzewskiInventsKPGroup, Lukierski_kappaPoincareanydimension} to quantize solvable Lie algebras employing them, which  is used  to obtain the quantum Poincar\'e\ groups associated with $\kappa$- and $\varrho$- Minkowski spacetimes.

Given a Lie algebra $\mathfrak{g}$ the  \textit{classical $r$-matrix} is 
a tensor $r\in \bigwedge^2 \mathfrak{g}$,
satisfying  the  MYBE, namely 
\begin{equation}
[[r,r]]= t \label{1.48}
\end{equation}
with $t\in  \otimes^3 \mathfrak{g}$ a $\mathfrak{g}$-invariant element and $[[r,r]]=[r_{12},r_{13}+ r_{23}]+[r_{13},r_{23}]$ . In the case $[[r,r]]=0$, this is the CYBE.
Here, $r_{\alpha \beta}\in \otimes^3 \mathfrak{g}$, $\alpha,\beta=1,2,3$ so that 
 $$
r_{12} =c_{ij} a_i \otimes a_j \otimes 1,\;\; r_{23} =c_{ij} 1 \otimes a_i \otimes a_j,\;\; r_{13} =c_{ij} a_i \otimes 1 \otimes a_j,
$$
with $a_i\in \mathfrak{g}$. 

The classical $r$-matrix has the important property of defining a Lie bialgebra structure on the Lie algebra $\mathfrak{g}$. Moreover, it allows for the definition of a Poisson bracket on the group manifold, which is compatible with the group structure, yielding to the notion of Poisson-Lie group, whose quantum counterpart is a quantum group. 
Hence, a Poisson-Lie group $G$ is  a  Lie group with group operations being Poisson maps~\cite{charipressley}. The algebra of smooth functions   $\mathcal{C}^\infty(G)$ is  a Hopf algebra (with trivial cosector and antipode), which is referred to as a  \textit{Poisson-Hopf algebra}. The  classical $r$-matrix provides the  Poisson-Lie structure through the following \textit{Sklyanin bracket}:
\begin{equation}
\{f,g\}=r^{\alpha \beta} (X_\alpha^RfX_\beta^Rg-X_\alpha^LfX_\beta^Lg), \hspace{1cm} f,g\in \mathcal{C}^\infty(G), \label{1.52}
\end{equation}
where $X^L$, $X^R$ are the left- and right-invariant vector fields.

If there are no ordering issues, one can quantize the  Poisson-Hopf algebra by means of  the usual canonical quantization $\{,\}\rightarrow \frac{1}{i}[,]$ to obtain the corresponding quantum Hopf algebra, namely, the quantum group. This is the case for both $\kappa$-Poincar\'e\ and $\varrho$- Poincar\'e\ deformations considered in \ref{rmatrixalgebra} and \ref{rhopoincare}.

\subsection*{Acknowledgments} We thank Giulia Gubitosi for useful discussions.
We acknowledge support from the INFN Iniziativa Specifica GeoSymQFT.  F.L.~acknowledges financial support from the State Agency for Research of the Spanish Ministry of Science
and Innovation through the ``Unit of Excellence Maria de Maeztu 2020--2023'' award to the Institute of Cosmos Sciences (Grant No. CEX2019-000918-M) and from Grants No. PID2019–105614 GB-C21 and No. 2017-SGR-929. L.S. acknowledges financial support from the Silicon Valley Community Foundation through the ``Agency-dependent spacetime and spacetime-dependent agency'' project.

\end{document}